\documentstyle[aps]{revtex}

\begin{document}
\title{Conserving approximations {\it vs} Two-Particle Self-Consistent
Approach}
\author{S. Allen$^{1,3}$, A.-M. S. Tremblay$^{1,2}$, and Y.M. Vilk$^{4}$,}
\address{$^{1}$D\'{e}partement de physique and Centre de recherche \\
sur les propri\'{e}t\'{e}s \'{e}lectroniques de mat\'{e}riaux avanc\'{e}s\\
$^{2}$Institut canadien de recherches avanc\'{e}es,\\
Universit\'{e} de Sherbrooke, Sherbrooke, Qu\'{e}bec, Canada J1K 2R1.\\
$^{3}$Centre de recherches math\'{e}matiques, Universit\'{e} de\\
Montr\'{e}al, Case postale 6128, Succursale centre-ville,\\
Montr\'{e}al, Qu\'{e}bec, Canada H3C 3J7\\
$^{4}$2100 Valencia Dr. apt. 406, Northbrook, IL 60062}
\date{\today}
\maketitle

\begin{abstract}
The conserving approximation scheme to many-body problems was developed by
Kadanoff and Baym using the functional-derivative approach. Another approach
for the Hubbard model also satisfies conservation laws, but in addition it
satisfies the Pauli principle and a number of sum rules. A concise formal
derivation of that approach, using functional derivatives, is given in this
conference paper to highlight formal analogies and differences with
conserving approximations.
\end{abstract}

\section{Introduction}

Although quantum many-body theory is in principle exact, any practical
application of the formalism necessarily leads to the violation of some
exact results. For example, it was noticed long ago that whenever infinite
subsets of diagrams are resummed, conservation laws will be violated if
certain rules are not followed. As another example of a possible
inconsistency, one can obtain a relation between number of particles,
chemical potential and temperature from either a Green function or from a
derivative of the free energy. How can we make sure that the diagrammatic
series for the Green function and that for the free energy give the same
result? Both of these problems, namely obtaining response functions that
satisfy conservation laws and making sure that the free energy is
\textquotedblleft thermodynamically consistent\textquotedblright\ with the
Green function, can be solved using an approach that was developed by
Kadanoff and Baym\cite{Baym:1962} \cite{BaymKadanoff:1962} many years ago.
This approach, described in the paper by Bickers (Chapter 6) in these
Proceedings, allows one to obtain so-called \textquotedblleft conserving
approximations\textquotedblright \cite{Bickers:1989} that have been
extensively used in the last decade. However, as in any approximate
calculation, some exact identities are not satisfied by conserving
approximations.\cite{Vilk:1997,Stamp:1985} For example, the Pauli principle
is violated, integration over coupling constant of the potential energy
obtained from susceptibilities does not lead back to the starting free
energy, and the irreducible verticies generated by a functional derivative
of the self-energy are not the same as the vertex corrections entering the
self-energy itself.\cite{NoteFLEX} In addition, since there is an infinite
number of choice of diagrams for the Luttinger-Ward functional, there is
also an infinite number of conserving approximations and it is difficult to
decide on the validity of an approximation based on the criterion of
conservation only. Further criticism of this approach has appeared in Refs.%
\cite{Vilk:1997} and \cite{Vilk:1996}

In this paper, we wish to present a new derivation of a {\it non-diagrammatic%
} approach to the {\it repulsive} Hubbard model that was developed a few
years ago.\cite{Vilk:1997} An analogous derivation in the case of the {\it %
attractive }Hubbard model was presented at this workshop by Bumsoo Kyung and
has already appeared.\cite{Allen:2001,Kyung:2001} Here, we go back to our
earlier results on the repulsive Hubbard model, but we use a formal approach
that highlights analogies and differences with conserving approximations.
This paper will strive to be concise with the explicit purpose of being
clear without getting lost into details. The published derivation for the
attractive Hubbard model contains more of these details.\cite{Allen:2001}

In section \ref{Section:conserving} we first present a derivation of
conserving approximations that can be extended, in section \ref{Section:Vilk}%
, to the derivation of our non-diagrammatic approach. This derivation
focuses first on response functions and on the self-consistent determination
of irreducible spin and charge vertices. That determination is done in such
a way that the Pauli principle in its simplest form is satisfied. In other
words, crossing symmetry (Chapter 6) is satisfied only for field operators
on a single site, all at the same time. This is the so-called two-particle
self-consistent (TPSC) part of the approach. We then improve our
approximation for the self-energy by using the TPSC results in the exact non
Hartree-Fock part of the self-energy. Vertices and Green function entering
the final expression are at the same level of approximation. No Migdal
theorem is assumed. The effects of transverse and longitudinal fluctuations
are considered in separate subsections before the final expression can be
written down. The internal-accuracy check\cite{Vilk:1997} is also presented
and used in the derivation. We end up in Sec. \ref{Section:Discussion} with
a discussion of extensions of the approach that are now being developed. The
appendix addresses a technical detail.

\section{Functional derivative formalism and conserving approximations}

\label{Section:conserving}

In this section, we first discuss single-particle properties, then response
functions and, finally, we present the Hartree-Fock approximation as an
example of a conserving approximation. The latter will be helpful as a
starting point for the derivation of the two-particle self-consistent (TPSC)
approach in the next section.

\subsection{Single-particle properties}

Following functional methods of the Schwinger school\cite%
{Baym:1962,BaymKadanoff:1962,MartinSchwinger:1959}, we begin with the
generating function with source fields $\phi _{\sigma }$ and field
destruction operators $\psi $ in the grand canonical ensemble 
\begin{equation}
\ln Z\left[ \phi \right] =\ln {\rm Tr\quad }[e^{-\beta \left( \widehat{H}%
-\mu \widehat{N}\right) }\text{T}_{\tau }\left( e^{-\psi _{\overline{\sigma }%
}^{\dagger }\left( \overline{1}\right) \phi _{\overline{\sigma }}\left( 
\overline{1},\overline{2}\right) \psi _{\overline{\sigma }}\left( \overline{2%
}\right) }\right) ]  \label{Generatrice}
\end{equation}
We adopt the convention that $1$ stands for the position and imaginary time
indices $\left( {\bf r}_{1},\tau _{1}\right) .$ The overbar means summation
over every lattice site and integration over imaginary-time from $0$ to $%
\beta $. T$_{\tau }$ is the time-ordering operator.

The propagator in the presence of the source field is obtained from
functional differentiation

\begin{equation}
G_{\sigma }\left( 1,2;\{\phi \}\right) =-\left\langle \psi _{\sigma }\left(
1\right) \psi _{\sigma }^{\dagger }\left( 2\right) \right\rangle _{\phi }=-%
\frac{\delta \ln Z\left[ \phi \right] }{\delta \phi _{\sigma }\left(
2,1\right) }.  \label{propagateur}
\end{equation}
From now on, {\it the time-ordering operator in averages, }$\left\langle
{}\right\rangle $, {\it is implicit}. Physically relevant correlation
functions are obtained for $\{\phi \}=0$ but it is extremely convenient to
keep finite $\{\phi \}$ in intermediate steps of the calculation.

Using the equation of motion for the field $\psi $ and the definition of the
self-energy, one obtains the Dyson equation in the presence of the source
field\cite{KB43} 
\begin{equation}
\left( G_{0}^{-1}-\phi \right) G=1+\Sigma G\quad ;\quad
G^{-1}=G_{0}^{-1}-\phi -\Sigma  \label{Dyson}
\end{equation}
where, from the commutator of the interacting part of the Hubbard
Hamiltonian $H,$ one obtains 
\begin{equation}
\Sigma _{\sigma }\left( 1,\overline{1};\{\phi \}\right) G_{\sigma }\left( 
\overline{1},2;\{\phi \}\right) =-U\left\langle \psi _{-\sigma }^{\dagger
}\left( 1^{+}\right) \psi _{-\sigma }\left( 1\right) \psi _{\sigma }\left(
1\right) \psi _{\sigma }^{\dagger }\left( 2\right) \right\rangle _{\phi }.
\label{Sigma*G}
\end{equation}
The imaginary time in $1^{+}$ is infinitesimally larger than in $1$.

\subsection{Response functions}

Response (four-point) functions for spin and charge excitations can be
obtained from functional derivatives $\left( \delta G/\delta \phi \right) $
of the source-dependent propagator. Following the standard approach and
using matrix notation to abbreviate the summations and integrations we have,

\begin{equation}
GG^{-1}=1  \label{GG-1=1}
\end{equation}
\begin{equation}
\frac{\delta G}{\delta \phi }G^{-1}+G\frac{\delta G^{-1}}{\delta \phi }=0.
\end{equation}
Using the Dyson equation (\ref{Dyson}) $G^{-1}=G_{0}^{-1}-\phi -\Sigma $
this may be rewritten

\begin{equation}
\frac{\delta G}{\delta \phi }=-G\frac{\delta G^{-1}}{\delta \phi }G=G_{%
\symbol{94}}G+G\frac{\delta \Sigma }{\delta \phi }G,  \label{RPA_phi}
\end{equation}
where the symbol $_{\symbol{94}}$ reminds us that the neighboring labels of
the propagators have to be the same as those of the $\phi $ in the
functional derivative. If perturbation theory converges, we may write the
self-energy as a functional of the propagator$.$ From the chain rule, one
then obtains an integral equation for the response function in the
particle-hole channel that is the analog of the Bethe-Salpeter equation in
the particle-particle channel 
\begin{equation}
\frac{\delta G}{\delta \phi }=G_{\symbol{94}}G+G\left[ \frac{\delta \Sigma }{%
\delta G}\frac{\delta G}{\delta \phi }\right] G.  \label{RPA_G}
\end{equation}
The labels of the propagators in the last term are attached to the self
energy, as in Eq.(\ref{RPA_phi}).\cite{NoteVertical} Vertices appropriate
for spin and charge responses are given, respectively, by 
\begin{equation}
U_{sp}=\frac{\delta \Sigma _{\uparrow }}{\delta G_{\downarrow }}-\frac{%
\delta \Sigma _{\uparrow }}{\delta G_{\uparrow }}\quad ;\quad U_{ch}=\frac{%
\delta \Sigma _{\uparrow }}{\delta G_{\downarrow }}+\frac{\delta \Sigma
_{\uparrow }}{\delta G_{\uparrow }}.  \label{dSigma/dG}
\end{equation}

\subsection{Hartree-Fock and RPA as an example}

As an example of calculation of response functions, consider the
Hartree-Fock approximation which corresponds to factoring the four-point
function in the definition of the self-energy Eq.(\ref{Sigma*G}) as if there
were no correlations,

\begin{equation}
\Sigma _{\sigma }^{H}\left( 1,\overline{1};\{\phi \}\right) G_{\sigma
}^{H}\left( \overline{1},2;\{\phi \}\right) =UG_{-\sigma }^{H}\left(
1,1^{+};\{\phi \}\right) G_{\sigma }^{H}\left( 1,2;\{\phi \}\right) .
\end{equation}%
Multiplying the above equation by $\left( G_{\sigma }^{H}\right) ^{-1},$ we
are left with 
\begin{equation}
\Sigma _{\sigma }^{H}\left( 1,2;\{\phi \}\right) =UG_{-\sigma }^{H}\left(
1,1^{+};\{\phi \}\right) \delta \left( 1-2\right) ,
\end{equation}%
\begin{equation}
\left. \frac{\delta \Sigma _{\uparrow }^{H}\left( 1,2;\{\phi \}\right) }{%
\delta G_{\downarrow }^{H}\left( 3,4;\{\phi \}\right) }\right\vert _{\{\phi
\}=0}=U\delta \left( 1-2\right) \delta \left( 3-1\right) \delta \left(
4-2\right) ,
\end{equation}%
which, when substituted in the integral equation (\ref{RPA_G}) for the
response function, tells us that we have generated the random phase
approximation (RPA) with, from Eq.(\ref{dSigma/dG}), $U_{sp}=U_{ch}=U.$

\section{Another approach}

\label{Section:Vilk}

The approach developed in Ref.\cite{Vilk:1997}, consists in two steps,
corresponding to the following two subsections. First, one computes response
functions from the TPSC approach, inspired by earlier work of Singwi.\cite%
{Singwi:1980} Second, one generates an improved approximation for the
self-energy starting from an exact expression for $\Sigma $ that explicitly
separates the infinite-frequency limit from the lower-frequency
contribution. The method also contains an internal-accuracy check that is
discussed in the last subsection.

\subsection{First step: two-particle self-consistency for $G^{\left(
1\right) },\Sigma ^{\left( 1\right) },$ $\Gamma _{sp}^{\left( 1\right)
}=U_{sp}$ and $\Gamma _{ch}^{\left( 1\right) }=U_{ch}$}

In conserving approximations, the self-energy is obtained from a functional
derivative $\Sigma \left[ G\right] =\delta \Phi \left[ G\right] /\delta G$
of $\Phi $ the Luttinger-Ward functional, which is itself computed from a
set of diagrams. To liberate ourselves from diagrams, we start instead from
the exact expression for the self-energy, Eq.(\ref{Sigma*G}) and notice that
when label $2$ equals $1^{+},$ the right-hand side of this equation is equal
to double-occupancy $\left\langle n_{\uparrow }n_{\downarrow }\right\rangle $%
. Factoring as in Hartree-Fock amounts to assuming no correlations. Instead,
we should insist that $\left\langle n_{\uparrow }n_{\downarrow
}\right\rangle $ should be obtained self-consistently. After all, in the
Hubbard model, there are only two local four point functions: $\left\langle
n_{\uparrow }n_{\downarrow }\right\rangle $ and $\left\langle n_{\uparrow
}^{2}\right\rangle =\left\langle n_{\downarrow }^{2}\right\rangle .$ The
latter is given exactly, through the Pauli principle, by $\left\langle
n_{\uparrow }^{2}\right\rangle =\left\langle n_{\downarrow
}^{2}\right\rangle =\left\langle n_{\uparrow }\right\rangle =\left\langle
n_{\downarrow }\right\rangle =n/2,$ when the filling $n$ is known$.$ In a
way, $\left\langle n_{\uparrow }n_{\downarrow }\right\rangle $ in the
self-energy equation (\ref{Sigma*G}), can be considered as an initial
condition for the four point function when one of the points, $2$, separates
from all the others which are at $1.$ When that label $2$ does not coincide
with $1$, it becomes more reasonable to factor {\it \`{a} la} Hartree-Fock.
These physical ideas are implemented by postulating

\begin{mathletters}
\begin{equation}
\Sigma _{\sigma }^{\left( 1\right) }\left( 1,\overline{1};\{\phi \}\right)
G_{\sigma }^{\left( 1\right) }\left( \overline{1},2;\{\phi \}\right)
=A_{\{\phi \}}G_{-\sigma }^{\left( 1\right) }\left( 1,1^{+};\{\phi \}\right)
G_{\sigma }^{\left( 1\right) }\left( 1,2;\{\phi \}\right)  \eqnum{13}
\label{ansatz}
\end{equation}
where $A_{\{\phi \}}$ depends on external field and is chosen such that the
exact result \cite{Note1+} 
\end{mathletters}
\begin{equation}
\Sigma _{\sigma }\left( 1,\overline{1};\{\phi \}\right) G_{\sigma }\left( 
\overline{1},1^{+};\{\phi \}\right) =U\left\langle n_{\uparrow }\left(
1\right) n_{\downarrow }\left( 1\right) \right\rangle _{\phi }
\label{SigmaG=un_upn_dwn}
\end{equation}
is satisfied. It is easy to see that the solution is 
\begin{equation}
A_{\{\phi \}}=U\frac{\left\langle n_{\uparrow }\left( 1\right) n_{\downarrow
}\left( 1\right) \right\rangle _{\phi }}{\left\langle n_{\uparrow }\left(
1\right) \right\rangle _{\phi }\left\langle n_{\downarrow }\left( 1\right)
\right\rangle _{\phi }}.
\end{equation}
Substituting $A_{\{\phi \}}$ back into our {\it ansatz} Eq.(\ref{ansatz}) we
obtain the self-energy by right-multiplying by $\left( G_{\sigma }^{\left(
1\right) }\right) ^{-1}:$%
\begin{equation}
\Sigma _{\sigma }^{\left( 1\right) }\left( 1,2;\{\phi \}\right) =A_{\{\phi
\}}G_{-\sigma }^{\left( 1\right) }\left( 1,1^{+};\{\phi \}\right) \delta
\left( 1-2\right) .
\end{equation}

We are now ready to obtain irreducible vertices using the prescription of
the previous section, Eq.(\ref{dSigma/dG}), namely through functional
derivatives of $\Sigma $ with respect to $G.$ In the calculation of $U_{sp},$
the functional derivative of $\left\langle n_{\uparrow }n_{\downarrow
}\right\rangle /\left( \left\langle n_{\uparrow }\right\rangle \left\langle
n_{\downarrow }\right\rangle \right) $ drops out, so we are left with,\cite%
{Note-ph} 
\begin{equation}
U_{sp}=\left. \frac{\delta \Sigma _{\uparrow }^{\left( 1\right) }}{\delta
G_{\downarrow }^{\left( 1\right) }}\right| _{\{\phi \}=0}-\left. \frac{%
\delta \Sigma _{\uparrow }^{\left( 1\right) }}{\delta G_{\uparrow }^{\left(
1\right) }}\right| _{\{\phi \}=0}=A_{\{\phi \}=0}=U\frac{\left\langle
n_{\uparrow }n_{\downarrow }\right\rangle }{\left\langle n_{\uparrow
}\right\rangle \left\langle n_{\downarrow }\right\rangle }.  \label{Usp}
\end{equation}
The renormalization of this irreducible vertex may be physically understood
as coming from Kanamori-Brueckner screening.\cite{Vilk:1997} To close the
system of equations, we need to know double-occupancy. It may be found
self-consistently using the fluctuation-dissipation theorem and{\bf \ }the
Pauli principle. First notice that the Pauli principle, $\left\langle
n_{\sigma }^{2}\right\rangle =\left\langle n_{\sigma }\right\rangle $,
implies that 
\begin{equation}
\left\langle \left( n_{\uparrow }-n_{\downarrow }\right) ^{2}\right\rangle
=\left\langle n_{\uparrow }\right\rangle +\left\langle n_{\downarrow
}\right\rangle -2\left\langle n_{\uparrow }n_{\downarrow }\right\rangle
\end{equation}
while the fluctuation-dissipation theorem tells us that $\left\langle \left(
n_{\uparrow }-n_{\downarrow }\right) ^{2}\right\rangle $ is given by the
equal-time equal-position imaginary-time susceptibility $\chi _{sp}.$ Since $%
\chi _{sp}^{-1}\left( q\right) =\chi _{0}^{-1}(q)-\frac{1}{2}U_{sp}$ with $%
q\equiv \left( {\bf q,}2\pi nT\right) $, this is equivalent to the equation 
\begin{equation}
\frac{T}{N}\sum_{q}\frac{\chi _{0}(q)}{1-\frac{1}{2}U_{sp}\chi _{0}(q)}%
=n-2\langle n_{\uparrow }n_{\downarrow }\rangle  \label{FDspin}
\end{equation}
that, with Eq.(\ref{Usp}) for $U_{sp}$, gives us double-occupancy.

The functional-derivative procedure generates an expression for the charge
vertex $U_{ch}$ which involves the functional derivative of $\left\langle
n_{\uparrow }n_{\downarrow }\right\rangle /\left( \left\langle n_{\uparrow
}\right\rangle \left\langle n_{\downarrow }\right\rangle \right) $ which
contains six point functions that one does not really know how to evaluate.
But, if we again assume that the vertex $U_{ch}$ is a constant, it is simply
determined by the requirement that charge fluctuations also satisfy the
fluctuation-dissipation theorem and the Pauli principle, 
\begin{equation}
\frac{T}{N}\sum_{q}\frac{\chi _{0}(q)}{1+\frac{1}{2}U_{ch}\chi _{0}(q)}%
=n+2\langle n_{\uparrow }n_{\downarrow }\rangle -n^{2}.  \label{FDcharge}
\end{equation}

Note that, in principle, $\Sigma ^{\left( 1\right) }$ also depends on
double-occupancy, but since $\Sigma ^{\left( 1\right) }$ is a constant, it
is absorbed in the definition of the chemical potential and we do not need
to worry about it in this case. That is why the non-interacting irreducible
susceptibility $\chi _{0}(q)$ appears in the expressions for the
susceptibility, even though it should be evaluated with $G^{\left( 1\right)
} $ that contains $\Sigma ^{\left( 1\right) }.$ One can check that spin and
charge conservation are satisfied by our susceptibilities.

Detailed comparisons with quantum Monte Carlo simulations (QMC)\cite%
{Chen:1994,Vilk:1997,Vilk:1996,Kyung:2003} have shown that Eqs.(\ref{Usp})(%
\ref{FDspin})(\ref{FDcharge}) give predictions that are quantitative at the
few percent level, in regions of parameter space where size effects are
negligible. This remains true even at couplings of the order of the
bandwidth and when second-neighbor hopping $t^{\prime }$ is present.\cite%
{Veilleux:1994} QMC size effects become important at half-filling below a
crossover temperature $T_{X}$ where the renormalized-classical regime
appears. Even though equation (\ref{Usp}) for $U_{sp}$ fails at $n=1$, $%
t^{\prime }=0$ in this regime, by assuming that $\langle n_{\uparrow
}n_{\downarrow }\rangle $ is temperature independent below $T_{X}$ one
obtains a qualitatively correct description of the renormalized-classical
regime. The universality class of our theory however is $O\left( N=\infty
\right) $ instead of $O\left( N=3\right) .$\cite{Dare:1996} A rough estimate
of the renormalized chemical potential (or equivalently of $\Sigma ^{\left(
1\right) }$), is given in appendix.

\subsection{Second step: an improved self-energy $\Sigma ^{\left( 2\right) }$%
}

Collective modes are less influenced by details of the single-particle
properties than the other way around. We thus wish now to obtain an improved
approximation for the self-energy that takes advantage of the fact that we
have found accurate approximations for the low-frequency spin and charge
fluctuations. We begin from the general definition of the self-energy Eq.(%
\ref{Sigma*G}) obtained from Dyson's equation. The right-hand side of that
equation can be obtained either from a functional derivative with respect to
an external field that is diagonal in spin, as in our generating function
Eq.(\ref{Generatrice}), or by a functional derivative of $\left\langle \psi
_{-\sigma }\left( 1\right) \psi _{\sigma }^{\dagger }\left( 2\right)
\right\rangle _{\phi _{t}}$ with respect to a transverse external field $%
\phi _{t}.$ These two approaches will be considered in turn below. They give
a self-energy formula that takes into account, respectively, longitudinal
and transverse fluctuations. Crossing symmetry, rotational symmetry and sum
rules will dictate the final formula for the improved self-energy $\Sigma
^{\left( 2\right) }$ that will be presented at the end of the subsection on
the consistency check.

\subsubsection{Longitudinal channel}

The right-hand side of the general definition of the self-energy Eq.(\ref%
{Sigma*G}) may be written as

\begin{equation}
\Sigma _{\sigma }\left( 1,\overline{1}\right) G_{\sigma }\left( \overline{1}%
,2\right) =-U\left[ \left. \frac{\delta G_{\sigma }\left( 1,2;\{\phi
\}\right) }{\delta \phi _{-\sigma }\left( 1^{+},1\right) }\right\vert
_{\{\phi \}=0}-G_{-\sigma }\left( 1,1^{+}\right) G_{\sigma }\left(
1,2\right) \right] .
\end{equation}%
The last term is the Hartree-Fock contribution. It gives the exact result
for the self-energy in the limit $\omega \rightarrow \infty $.\cite%
{Vilk:1997} The $\delta G_{\sigma }/\delta \phi _{-\sigma }$ term is thus a
contribution to lower frequencies and it comes from the spin and charge
fluctuations. Right-multiplying the last equation by $G^{-1}$ and replacing
the lower energy part $\delta G_{\sigma }/\delta \phi _{-\sigma }$ by its
general expression in terms of irreducible vertices, Eq.(\ref{RPA_G}) we
find 
\begin{equation}
\Sigma _{\sigma }^{\left( 2\right) }\left( 1,2\right) =UG_{-\sigma }^{\left(
1\right) }\left( 1,1^{+}\right) \delta \left( 1-2\right) -UG_{\sigma
}^{\left( 1\right) }\left( 1,\overline{3}\right) \left[ \left. \frac{\delta
\Sigma _{\sigma }^{\left( 1\right) }\left( \overline{3},2;\{\phi \}\right) }{%
\delta G_{\overline{\sigma }}^{\left( 1\right) }\left( \overline{4},%
\overline{5};\{\phi \}\right) }\right\vert _{\{\phi \}=0}\left. \frac{\delta
G_{\overline{\sigma }}^{\left( 1\right) }\left( \overline{4},\overline{5}%
;\{\phi \}\right) }{\delta \phi _{-\sigma }\left( 1^{+},1\right) }%
\right\vert _{\{\phi \}=0}\right] .  \label{SigmaLongExact}
\end{equation}%
Every quantity appearing on the right-hand side of that equation has been
taken from the TPSC results. This means in particular that the irreducible
vertices $\delta \Sigma _{\sigma }^{\left( 1\right) }/\delta G_{\sigma
^{\prime }}^{\left( 1\right) }$ are at the same level of approximation as
the Green functions $G_{\sigma }^{\left( 1\right) }$ and self-energies $%
\Sigma _{\sigma }^{\left( 1\right) }.$ In approaches that assume that
Migdal's theorem applies to spin and charge fluctuations, one often sees
renormalized Green functions $G^{\left( 2\right) }$ appearing on the
right-hand side along with unrenormalized vertices, $\delta \Sigma _{\sigma
}/\delta G_{\sigma ^{\prime }}\rightarrow U.$

In terms of $U_{sp}$ and $U_{ch}$ in Fourier space, the above formula\cite%
{Vilk:1996} reads,

\begin{equation}
\Sigma _{\sigma }^{\left( 2\right) }(k)_{long}=Un_{-\sigma }+\frac{U}{4}%
\frac{T}{N}\sum_{q}\left[ U_{sp}\chi _{sp}^{\left( 1\right) }(q)+U_{ch}\chi
_{ch}^{\left( 1\right) }(q)\right] G_{\sigma }^{\left( 1\right) }(k+q).
\label{Self-longitudinal}
\end{equation}

\subsubsection{Transverse channel}

In the transverse channel, the calculation basically has to be redone from
scratch. It is closely analogous to the attractive Hubbard model case, for
which detailed calculations have been published.\cite{Allen:2001} We will
thus be concise. The generating function in a transverse field is

\begin{equation}
\ln Z\left[ \phi _{t}\right] =\ln {\rm Tr\quad }[e^{-\beta \left( \widehat{H}%
-\mu \widehat{N}\right) }\text{T}_{\tau }\left( e^{-\psi _{\uparrow
}^{\dagger }\left( \overline{1}\right) \phi _{-}\left( \overline{1},%
\overline{2}\right) \psi _{\downarrow }\left( \overline{2}\right) -\psi
_{\downarrow }^{\dagger }\left( \overline{1}\right) \phi _{+}\left( 
\overline{1},\overline{2}\right) \psi _{\uparrow }\left( \overline{2}\right)
}\right) ]  \label{Generatrice_transv}
\end{equation}
The corresponding spin-space matrix Green function 
\begin{equation}
{\bf G}\left( 1,2;\left\{ \phi _{t}\right\} \right) =-\left( 
\begin{array}{cc}
\left\langle \psi _{\uparrow }\left( 1\right) \psi _{\uparrow }^{\dagger
}\left( 2\right) \right\rangle _{\phi _{t}} & \left\langle \psi _{\uparrow
}\left( 1\right) \psi _{\downarrow }^{\dagger }\left( 2\right) \right\rangle
_{\phi _{t}} \\ 
\left\langle \psi _{\downarrow }\left( 1\right) \psi _{\uparrow }^{\dagger
}\left( 2\right) \right\rangle _{\phi _{t}} & \left\langle \psi _{\downarrow
}\left( 1\right) \psi _{\downarrow }^{\dagger }\left( 2\right) \right\rangle
_{\phi _{t}}%
\end{array}
\right)
\end{equation}
obeys the matrix Dyson equation 
\begin{equation}
{\bf G}^{-1}\left( 1,2;\left\{ \phi _{t}\right\} \right) ={\bf G}%
_{0}^{-1}\left( 1-2\right) -{\bf \Sigma }\left( 1,2;\left\{ \phi
_{t}\right\} \right) -\Phi _{t}
\end{equation}
with 
\begin{equation}
\Phi _{t}=\left( 
\begin{array}{cc}
0 & \phi _{-} \\ 
\phi _{+} & 0%
\end{array}
\right)
\end{equation}
and 
\begin{equation}
{\bf \Sigma }\left( 1,\overline{1};\{\phi _{t}\}\right) {\bf G}\left( 
\overline{1},2;\{\phi _{t}\}\right) \equiv -U\left( 
\begin{array}{cc}
\left\langle \psi _{\downarrow }^{\dagger }\left( 1^{+}\right) \psi
_{\downarrow }\left( 1\right) \psi _{\uparrow }\left( 1\right) \psi
_{\uparrow }^{\dagger }\left( 2\right) \right\rangle _{\phi _{t}} & 
\left\langle \psi _{\downarrow }^{\dagger }\left( 1^{+}\right) \psi
_{\downarrow }\left( 1\right) \psi _{\uparrow }\left( 1\right) \psi
_{\downarrow }^{\dagger }\left( 2\right) \right\rangle _{\phi _{t}} \\ 
\left\langle \psi _{\uparrow }^{\dagger }\left( 1^{+}\right) \psi _{\uparrow
}\left( 1\right) \psi _{\downarrow }\left( 1\right) \psi _{\uparrow
}^{\dagger }\left( 2\right) \right\rangle _{\phi _{t}} & \left\langle \psi
_{\uparrow }^{\dagger }\left( 1^{+}\right) \psi _{\uparrow }\left( 1\right)
\psi _{\downarrow }\left( 1\right) \psi _{\downarrow }^{\dagger }\left(
2\right) \right\rangle _{\phi _{t}}%
\end{array}
\right)  \label{Sigma*G_t}
\end{equation}

We need to obtain, in turn, the renormalized vertex $U_{sp}$ Eq.(\ref{Usp}),
and finally find an improved formula for the self-energy $\Sigma ^{\left(
2\right) }$. Starting from the matrix equation ${\bf GG}^{-1}={\bf 1}$ and
following the procedure below Eq.(\ref{GG-1=1}), one of the two response
functions that do not vanish in zero external field obeys, 
\begin{equation}
\left. \frac{\delta G_{21}\left( 1,2;\{\phi _{t}\}\right) }{\delta \phi
_{+}\left( 3,4\right) }\right| _{\{\phi _{t}\}=0}=G_{22}\left( 1,3\right)
G_{11}\left( 4,2\right) +G_{22}\left( 1,\overline{2}\right) \left. \frac{%
\delta \Sigma _{21}\left( \overline{2},\overline{3};\{\phi _{t}\}\right) }{%
\delta G_{21}\left( \overline{6},\overline{7};\{\phi _{t}\}\right) }\right|
_{\{\phi _{t}\}=0}\left. \frac{\delta G_{21}\left( \overline{6},\overline{7}%
;\{\phi _{t}\}\right) }{\delta \phi _{+}\left( 3,4\right) }\right| _{\{\phi
_{t}\}=0}G_{11}\left( \overline{3},2\right)  \label{Reponse_transverse}
\end{equation}
where the subscripts denote matrix elements in spin-space. There is an
analogous equation for $\delta G_{12}\left( 1,2\right) /\delta \phi
_{-}\left( 3,4\right) .$ Note that 
\begin{equation}
\delta G_{21}\left( 1,1\right) /\delta \phi _{+}\left( 2,2\right)
=-\left\langle S_{+}\left( 1\right) S_{-}\left( 2\right) \right\rangle
=-\left\langle \psi _{\uparrow }^{\dagger }\left( 1\right) \psi _{\downarrow
}\left( 1\right) \psi _{\downarrow }^{\dagger }\left( 2\right) \psi
_{\uparrow }\left( 2\right) \right\rangle =-\chi _{+-}\left( 1,2\right) .
\end{equation}
The usual Hartree-Fock factorization for the self-energy would transform Eq.(%
\ref{Reponse_transverse}) into the RPA equation for transverse spin
fluctuations $\left\langle S_{+}\left( 1\right) S_{-}\left( 2\right)
\right\rangle $.

The value of $\Sigma ^{\left( 1\right) }$ and of the corresponding vertices
at the TPSC level are obtained using steps analogous to those in the
longitudinal channel. We factor the self-energy in the Hartree-Fock manner, 
\begin{equation}
{\bf \Sigma }^{\left( 1\right) }\left( 1,\overline{1};\{\phi _{t}\}\right) 
{\bf G}^{\left( 1\right) }\left( \overline{1},2;\{\phi _{t}\}\right)
=-A_{\{\phi _{t}\}}\left( 
\begin{array}{cc}
G_{22}^{\left( 1\right) }\left( 1,1^{+};\{\phi _{t}\}\right) & 
-G_{12}^{\left( 1\right) }\left( 1,1^{+};\{\phi _{t}\}\right) \\ 
-G_{21}^{\left( 1\right) }\left( 1,1^{+};\{\phi _{t}\}\right) & 
G_{11}^{\left( 1\right) }\left( 1,1^{+};\{\phi _{t}\}\right)%
\end{array}
\right) \left( 
\begin{array}{cc}
G_{11}^{\left( 1\right) }\left( 1,2;\{\phi _{t}\}\right) & G_{12}^{\left(
1\right) }\left( 1,2;\{\phi _{t}\}\right) \\ 
G_{21}^{\left( 1\right) }\left( 1,2;\{\phi _{t}\}\right) & G_{22}^{\left(
1\right) }\left( 1,2;\{\phi _{t}\}\right)%
\end{array}
\right) .  \label{HF_transv}
\end{equation}
but we correct this factorization by the factor $A_{\{\phi _{t}\}}$ which is
determined in such a way that when $2\rightarrow 1^{+}$ the exact result for
the four point function on the right-hand side of Eq.(\ref{Sigma*G_t}) is
recovered. Using $\left\langle \psi _{\downarrow }^{\dagger }\left(
1^{+}\right) \psi _{\downarrow }\left( 1\right) \psi _{\uparrow }\left(
1\right) \psi _{\downarrow }^{\dagger }\left( 1^{+}\right) \right\rangle
_{\phi _{t}}=\left\langle \psi _{\downarrow }^{\dagger }\left( 1^{+}\right)
\psi _{\downarrow }^{\dagger }\left( 1^{+}\right) \psi _{\downarrow }\left(
1\right) \psi _{\uparrow }\left( 1\right) \right\rangle _{\phi _{t}}=0$ and
the analogous result for the other off-diagonal element, the exact result
for $2\rightarrow 1^{+}$ is, 
\begin{equation}
{\bf \Sigma }^{\left( 1\right) }\left( 1,\overline{1};\{\phi _{t}\}\right) 
{\bf G}^{\left( 1\right) }\left( \overline{1},1^{+};\{\phi _{t}\}\right)
=U\left( 
\begin{array}{cc}
\left\langle n_{\downarrow }\left( 1\right) n_{\uparrow }\left( 1\right)
\right\rangle _{\phi _{t}} & 0 \\ 
0 & \left\langle n_{\uparrow }\left( 1\right) n_{\downarrow }\left( 1\right)
\right\rangle _{\phi _{t}}%
\end{array}
\right) .
\end{equation}
Equating with the right-hand side of the approximate result Eq.(\ref%
{HF_transv}) when $2\rightarrow 1^{+}$ determines the value of $A_{\{\phi
_{t}\}}$ by a simple $2\times 2$ matrix inversion. From this, one extracts
the off-diagonal component of the self energy and the corresponding
irreducible vertex that is needed for response functions in Eq.(\ref%
{Reponse_transverse}) 
\begin{equation}
\left. \frac{\delta \Sigma _{21}^{\left( 1\right) }\left( 1,2;\{\phi
_{t}\}\right) }{\delta G_{21}^{\left( 1\right) }\left( 3,4;\{\phi
_{t}\}\right) }\right| _{\{\phi _{t}\}=0}=-U\frac{\left\langle n_{\uparrow
}n_{\downarrow }\right\rangle }{\left\langle n_{\uparrow }\right\rangle
\left\langle n_{\downarrow }\right\rangle }\delta \left( 1-3\right) \delta
\left( 1-2\right) \delta \left( 1-4\right) .
\end{equation}
Substituting in the equation for the response function Eq.(\ref%
{Reponse_transverse}) one precisely recovers for the transverse spin
fluctuations the same TPSC result as for the longitudinal fluctuations: $%
\chi _{sp}^{-1}\left( q\right) =\chi _{0}^{-1}(q)-\frac{1}{2}U_{sp}$ with $%
U_{sp}$ given by Eq.(\ref{Usp}). The determination of $U_{sp}$ through the
fluctuation-dissipation theorem leads again to Eq.(\ref{FDspin}). We thus
recover the results expected from rotational invariance.

To move to the second level of approximation for the self-energy, we return
to the exact definition for the self-energy Eq.(\ref{Sigma*G_t}) but this
time in zero applied external field. The first diagonal component can be
written as 
\begin{equation}
\Sigma _{11}\left( 1,2\right) =U\left. \frac{\delta G_{21}\left( 1,\overline{%
2};\{\phi _{t}\}\right) }{\delta \phi _{+}\left( 1^{+},1\right) }\right\vert
_{\{\phi _{t}\}=0}G_{11}^{-1}\left( \overline{2},2\right) .
\end{equation}%
Using the exact result for the transverse response, Eq.(\ref%
{Reponse_transverse}) this takes the form 
\begin{equation}
\Sigma _{11}\left( 1,2\right) =UG_{22}\left( 1,1^{+}\right) \delta \left(
1-2\right) +UG_{22}\left( 1,\overline{3}\right) \left. \frac{\delta \Sigma
_{21}\left( \overline{3},2;\{\phi _{t}\}\right) }{\delta G_{21}\left( 
\overline{6},\overline{7};\{\phi _{t}\}\right) }\right\vert _{\{\phi
_{t}\}=0}\left. \frac{\delta G_{21}\left( \overline{6},\overline{7};\{\phi
_{t}\}\right) }{\delta \phi _{+}\left( 1^{+},1\right) }\right\vert _{\{\phi
_{t}\}=0}  \label{SigmaTrans}
\end{equation}%
where, as in the longitudinal case, the high-frequency Hartree-Fock result
is now explicit. Substituting on the right-hand side the TPSC ({\it i.e.}
level $\left( 1\right) $) results, we obtain an improved approximation for
the self-energy due to transverse spin fluctuations. In momentum space, it
reads,

\begin{equation}
\Sigma _{\sigma }^{\left( 2\right) }(k)_{trans}=Un_{-\sigma }+\frac{U}{2}%
\frac{T}{N}\sum_{q}\left[ U_{sp}\chi _{sp}^{\left( 1\right) }(q)\right]
G_{-\sigma }^{\left( 1\right) }(k+q).  \label{Self-transverse}
\end{equation}

\subsection{Internal accuracy check}

The equation (\ref{SigmaG=un_upn_dwn}) that relates ${\rm Tr}\left[ \Sigma G%
\right] $ to double occupancy (potential energy) relates purely
single-particle quantities, $\Sigma $ and $G,$ to a quantity that may be
computed from two-particle quantities, namely spin and charge correlation
functions. In our approach, it can be checked\cite{Vilk:1997} that $\frac{1}{%
2}{\rm Tr}\left[ \Sigma ^{\left( 2\right) }G^{\left( 1\right) }\right] $ is
exactly equal to $U\left\langle n_{\uparrow }n_{\downarrow }\right\rangle ,$
with $\left\langle n_{\uparrow }n_{\downarrow }\right\rangle $ computed at
the TPSC level. More specifically, we find,

\begin{equation}
\frac{1}{2}{\rm Tr}\left[ \Sigma ^{\left( 2\right) }G^{\left( 1\right) }%
\right] =\lim_{\tau \rightarrow 0^{-}}\frac{T}{N}\sum_{k}\Sigma _{\sigma
}^{\left( 2\right) }\left( k\right) G_{\sigma }^{\left( 1\right) }\left(
k\right) e^{-ik_{n}\tau }=U\left\langle n_{\uparrow }n_{\downarrow
}\right\rangle .  \label{Tr(sigma_G)}
\end{equation}%
One can use the difference between $\frac{1}{2}{\rm Tr}\left[ \Sigma
^{\left( 2\right) }G^{\left( 1\right) }\right] $ and $\frac{1}{2}{\rm Tr}%
\left[ \Sigma ^{\left( 2\right) }G^{\left( 2\right) }\right] $ as an
internal accuracy check of the theory. In the pseudogap regime at $n=1$ for
example, the breakdown of the theory is clearly indicated by the growing
difference between the two quantities. Ref.\cite{Kyung:2001} gives a
detailed table that illustrates these facts in the case of the attractive
Hubbard model.

The ${\rm Tr}\left[ \Sigma ^{\left( 2\right) }G^{\left( 1\right) }\right] $
formula can also be used to help in the interpretation of the two different
results obtained above for the self-energy, Eqs.(\ref{Self-longitudinal})
and (\ref{Self-transverse}). Without the approximation that the TPSC results
should be used on the right-hand side, the results that follow from the
corresponding exact expressions Eqs.(\ref{SigmaLongExact}) and (\ref%
{SigmaTrans}) would be identical. To resolve this problem, we follow Ref.%
\cite{Moukouri:2000}. Fig. 1 of this paper shows the self-energy in terms of
the fully reducible vertex $\Gamma \left( q,k-k^{\prime },k+k^{\prime
}-q\right) .$ In both formulas for the self-energy, Eqs.(\ref%
{Self-longitudinal}) and (\ref{Self-transverse}), the dependence of $\Gamma $
on the particle-particle channel center of mass momentum $k+k^{\prime }-q$
is neglected since this channel is not singular. The longitudinal version of
the self-energy Eq.(\ref{Self-longitudinal}) takes good care of the
singularity of $\Gamma $ when its first argument $q$ is near $\left( \pi
,\pi \right) .$ The transverse version does the same for the singular
dependence near $\left( \pi ,\pi \right) $ of the second argument $%
k-k^{\prime }$, which corresponds to the other particle-hole channel. One
then expects that averaging the two possibilities gives a better
approximation for $\Gamma $ since it preserves crossing symmetry in the two
particle-hole channels. Furthermore, one can verify that the longitudinal
spin fluctuations in Eq.(\ref{Self-longitudinal}) contribute an amount $%
U\left\langle n_{\uparrow }n_{\downarrow }\right\rangle /2$ to the
consistency condition $\frac{1}{2}{\rm Tr}\left( \Sigma _{long}^{\left(
2\right) }G^{\left( 1\right) }\right) =$ $U\left\langle n_{\uparrow
}n_{\downarrow }\right\rangle $ and that each of the two transverse spin
components also contribute $U\left\langle n_{\uparrow }n_{\downarrow
}\right\rangle /2$ to $\frac{1}{2}{\rm Tr}\left( \Sigma _{trans}^{\left(
2\right) }G^{\left( 1\right) }\right) =$ $U\left\langle n_{\uparrow
}n_{\downarrow }\right\rangle $ $.$ Hence, averaging Eq.(\ref%
{Self-longitudinal}) and the expression in the transverse channel Eq.(\ref%
{Self-transverse}) also preserves rotational invariance. In addition, one
verifies numerically that the exact sum rule\cite{Vilk:1997} $-\int d\omega
^{\prime }%
\mathop{\rm Im}%
\left[ \Sigma _{\sigma }\left( {\bf k,}\omega ^{\prime }\right) \right] /\pi
=U^{2}n_{-\sigma }\left( 1-n_{-\sigma }\right) $ determining the
high-frequency behavior is satisfied to a higher degree of accuracy. We thus
obtain a self-energy formula that we called\cite{Moukouri:2000}
\textquotedblleft symmetric\textquotedblright\ 
\begin{equation}
\Sigma _{\sigma }^{\left( 2\right) }(k)_{sym}=Un_{-\sigma }+\frac{U}{8}\frac{%
T}{N}\sum_{q}\left[ 3U_{sp}\chi _{sp}^{\left( 1\right) }(q)+U_{ch}\chi
_{ch}^{\left( 1\right) }(q)\right] G_{\sigma }^{\left( 1\right) }(k+q).
\label{Self-final}
\end{equation}%
$\Sigma _{\sigma }^{\left( 2\right) }(k)_{sym}$ is different from so-called
Berk-Schrieffer type expressions\cite{Berk:1966} that do not satisfy\cite%
{Vilk:1997} the consistency condition between one- and two-particle
properties, $\frac{1}{2}{\rm Tr}\left( \Sigma G\right) =$ $U\left\langle
n_{\uparrow }n_{\downarrow }\right\rangle .$

\section{Discussion and extensions}

\label{Section:Discussion}

The approach described above is valid in the weak to intermediate coupling
regime. It involves two steps. First, a self-consistent determination of
double-occupancy and renormalized vertices that enter the dynamical
susceptibilities of the most important four-point correlation functions
(density-density and spin-spin correlation functions). This is summarized by
Eqs.(\ref{Usp})(\ref{FDspin})(\ref{FDcharge}). Conservation laws, such as
charge and spin conservation, are satisfied. These results are then used to
obtain an improved approximation for single-particle properties through the
self-energy, Eq.(\ref{Self-final}). An internal accuracy check allows one to
decide on the validity of the results in cases where QMC or other exact
results are not available as references. The approach satisfies the Pauli
principle, the Mermin-Wagner theorem, contains Kanamori-Brueckner screening
and does not assume a Migdal theorem in the calculation of the self-energy.

In addition to provide an accurate calculational tool for the Hubbard model,
this methodology has allowed to develop insight into the Physics of this
model. The physically most important result obtained to date is probably the
detailed description of the Physics of the pseudogap that appears {\it in
two dimensions} in the renormalized-classical regime $\left( \hbar \omega
\ll k_{B}T\right) $ when the antiferromagnetic correlation length
(superconducting correlation length in the case of the attractive model)
becomes larger than the single particle thermal de Broglie wavelength. \cite%
{Vilk:1996,Vilk:1997} This Physics explains a possible route to the
destruction of the Fermi liquid in {\it two dimensions}. These results have
been confirmed by extensive QMC calculations.\cite%
{Allen:1999,Moukouri:2000,Kyung:2003} The pseudogap (depletion near $\omega
=0$) in $%
\mathop{\rm Im}%
G_{\sigma }^{R}\left( {\bf k}_{F},\omega \right) $ appears along with
precursors of the Bogoliubov quasiparticles (finite $\omega $ peaks) of the
ordered state. To extrapolate more deeply in the pseudogap regime, where
strictly speaking the above method fails, one assumes that double-occupancy
becomes temperature independent below the crossover temperature $T_{X}$
where one enters the renormalized-classical regime.

The above methodology has been applied successfully to the attractive
Hubbard model.\cite{Allen:2001,Kyung:2001,AllenTh:2000} It trivially applies
to the Hubbard model with an arbitrary hopping matrix. $\Sigma ^{\left(
2\right) }$ can also be used to obtain consistent thermodynamic predictions. %
\cite{Roy:2001} Extensions to multiband problems are non-trivial but are
being developed.\cite{Dare:2001} Recently, an extension of this approach has
been used to obtain quantitative results for the spin and charge
susceptibilities in the attractive Hubbard model.\cite{Kyung-cross:2001}
Proceeding along the same lines for the repulsive Hubbard model, pairing
correlations have been calculated.\cite{KyungLandry:2001} They yield a dome
shape dependence on doping of the $d-$wave superconducting transition
temperature $T_{c}$. The decrease of $T_{c}$ near half-filling comes from
the detrimental effect of opening a pseudogap. Phenomenological extensions
to more complicated models with {\it d}-wave superconductivity and
antiferromagnetism have also been proposed.\cite{Kyung-d-wave:2000} Future
directions include generalizations to the pseudogap regime, to states with
broken symmetry,\cite{Allen:2000} to longer range interactions and to
impurity models. It would also be extremely valuable to obtain the
irreducible vertices $\delta \Sigma _{\sigma }^{\left( 2\right) }/\delta
G_{\sigma ^{\prime }}^{\left( 2\right) }$ that are consistent with the best
estimate of the self-energy.

This paper is based in part on a course given by A.-M.S.T at Universit\'{e}
de Provence in 1999, and on seminars at the Newton Institute for
Mathematical Sciences and at the Institute for Theoretical Physics in Santa
Barbara in 2000. Work there was partially supported by the National Science
Foundation under grant No. PHY94-07194. A.-M.S.T. and S.Allen are grateful
to B. Kyung, F. Lemay and A.-M. Dar\'{e} for numerous discussions. A.-M.S.T.
thanks Gilbert Albinet and his group for hospitality. This work was
supported by a grant from the Natural Sciences and Engineering Research
Council (NSERC) of Canada and the Fonds pour la formation de Chercheurs et
l'Aide \`{a} la Recherche (FCAR) of the Qu\'{e}bec government. We thank the
Centre de recherches math\'{e}matiques for its hospitality. A.-M.S.T. holds
a Tier I Canada Research Chair in Condensed Matter Physics.

\appendix 

\section{An approximate formula for $\Sigma ^{\left( 1\right) }$}

The equation Eq.(\ref{SigmaG=un_upn_dwn}) that was used to obtain finally $%
U_{sp}=U\left\langle n_{\uparrow }n_{\downarrow }\right\rangle /\left(
\left\langle n_{\downarrow }\right\rangle \left\langle n_{\uparrow
}\right\rangle \right) $ would be different if we had taken the limit $%
2\rightarrow 1^{-}$ instead of $2\rightarrow 1^{+}$ in the general
expression for $\Sigma G$, Eq.(\ref{Sigma*G}). More specifically, at $\{\phi
\}=0$, 
\begin{equation}
\Sigma _{\sigma }\left( 1,\overline{1}\right) G_{\sigma }\left( \overline{1}%
,1^{+}\right) =U\left\langle n_{\downarrow }\left( 1\right) n_{\uparrow
}\left( 1\right) \right\rangle \quad ;\quad \Sigma _{\sigma }\left( 1,%
\overline{1}\right) G_{\sigma }\left( \overline{1},1^{-}\right)
=U\left\langle n_{\downarrow }\left( 1\right) \left( n_{\uparrow }\left(
1\right) -1\right) \right\rangle .  \label{Deux_possibilites}
\end{equation}
Any approximation for the self-energy that has Hartree-Fock as its
infinite-frequency limit will be such that the difference between the above
two results, $\Sigma _{\sigma }\left( 1,\overline{1}\right) G_{\sigma
}\left( \overline{1},1^{+}\right) -\Sigma _{\sigma }\left( 1,\overline{1}%
\right) G_{\sigma }\left( \overline{1},1^{-}\right) =U\left\langle
n_{\downarrow }\left( 1\right) \right\rangle $ is satisfied. The proof,
Eq.(44) of Ref.\cite{Allen:2001}, is as follows: 
\[
\frac{T}{N}\sum_{{\bf k}}\sum_{ik_{n}}\left( \frac{\Sigma \left( {\bf k}%
,ik_{n}\right) }{ik_{n}-\left( \varepsilon _{{\bf k}}-\mu \right) -\Sigma
\left( {\bf k},ik_{n}\right) }-\frac{U\left\langle n_{\downarrow
}\right\rangle }{ik_{n}}\right) \left(
e^{-ik_{n}0^{-}}-e^{-ik_{n}0^{+}}\right) 
\]
\begin{equation}
+\frac{T}{N}\sum_{{\bf k}}\sum_{ik_{n}}\left[ \frac{U\left\langle
n_{\downarrow }\right\rangle }{ik_{n}}\left(
e^{-ik_{n}0^{-}}-e^{-ik_{n}0^{+}}\right) \right] =U\left\langle
n_{\downarrow }\right\rangle .  \label{SigmaG_et_convergence}
\end{equation}
In this expression, the first sum vanishes because we have added and
subtracted a term that makes it convergent at infinity without the need for
convergence factors $e^{-ik_{n}0^{\pm }}$. Hence, only the last sum survives.

Since $\Sigma _{\sigma }^{\left( 1\right) }\left( 1,2\right) $ is a constant
times $\delta \left( 1-2\right) ,$ one can obtain two estimates of the
constant depending on which of the two equations in (\ref{Deux_possibilites}%
) one starts from. By analogy with Eq.(\ref{SigmaG_et_convergence}) one
expects that the difference between these two estimates is related to the
high frequency behavior of the true result while the average of the two
estimates is related to the low-frequency behavior. This average is 
\begin{equation}
\frac{n}{2}(U+U_{sp}\left( 1-n\right) )\frac{1}{2-n}.
\end{equation}
The exact result, $U/2$, is recovered at $n=1.$ For other fillings, the
above formula gives a very rough estimate of the chemical potential shift
induced by interactions. For example, for $U$ up to $4t$ on $8\times 8$
lattices and temperatures of order $t/4$ in energy units, one finds results
that deviate by up to $20\%$ from the QMC results. The corresponding
procedure in the {\it attractive} Hubbard model case\cite{Allen:2001} seems
to work better.


\begin{references}
\bibitem{NoteFLEX} In FLEX for example, the structure of the self-energy is
the same as if Migdal's theorem was applicable.
\bibitem{NoteVertical} To remind ourselves of this, we may also adopt an
additional ``vertical matrix notation'' convention and write Eq.(\ref%
{RPA_phi}) as $\frac{\delta G}{\delta \phi }=G_{\symbol{94}}G+G\left[ \frac{%
\frac{\delta \Sigma }{\delta G}}{\frac{\delta G}{\delta \phi }}\right] G.$
\bibitem{Note1+} If we had chosen $2\rightarrow 1^{-}$ instead of $%
2\rightarrow 1^{+}$ we would have found $\left\langle n_{\downarrow }\left(
1\right) \left( n_{\uparrow }\left( 1\right) -1\right) \right\rangle $ on
the right-hand side of Eq.(\ref{SigmaG=un_upn_dwn}), as discussed in the
appendix. Since we are interested in using the local limit of $\Sigma G$ as
an initial condition for spin and charge correlations, it is more natural to
focus on $2\rightarrow 1^{+}$ since in that case it is $\left\langle
n_{\downarrow }\left( 1\right) n_{\uparrow }\left( 1\right) \right\rangle $
that appears on the right-hand side, a quantity that follows directly from
the spin and charge correlations and where spin up and spin down are treated
on equal footing. See also the appendix.
\bibitem{Note-ph} For $n>1,$ all particle occupation numbers must be
replaced by hole occupation numbers.
\end{references}
\end{document}